\documentstyle[12pt,psfig,aasms4]{article}

\lefthead{Rosenberg A., et~al.}
\righthead{Galactic Globular Cluster Relative Ages.}

\begin{document}

\title{Galactic Globular Cluster Relative Ages.
\footnote{Based on observations collected at the European Southern
	Observatory, (La Silla, Chile), and the Isaac Newton Group of
	Telescopes, Observatorio del Roque de los Muchachos, (La Palma,
	Spain)} 
	}

\author{A. Rosenberg}
\affil{Telescopio Nazionale Galileo, Osservatorio Astronomico di Padova,
Italy; \\ rosenberg@pd.astro.it}
\author{I. Saviane, G. Piotto}
\affil{Dipartimento di Astronomia, Universit\`a di Padova, Italy; \\ saviane,piotto@pd.astro.it}
\author{A. Aparicio}
\affil{Instituto de Astrof\'{\i}sica de Canarias, Spain; \\ aaj@iac.es}

\begin{abstract}	
\label{abstract}

Based on a new large, homogeneous photometric database of 34 Galactic
globular clusters (+ Pal~12), a set of distance and reddening
independent relative age indicators has been measured. The observed
$\delta (V-I)_{@2.5}$ and $\Delta V^{\rm HB}_{\rm TO}$ vs. metallicity
relations have been compared to the relations predicted by two recent
updated libraries of isochrones. Using these models and two
independent methods, we have found that self-consistent relative ages
can be estimated for our GGC sample. In turn, this demonstrates that
the models are internally self-consistent.

Based on the relative age vs. metallicity distribution, we conclude
that: (a) there is no evidence of an age spread for clusters with
[Fe/H]$<-1.2$, all the clusters of our sample in this range being old and
coeval; (b) for the intermediate metallicity group
($-1.2\leq$[Fe/H]$<-0.9$) there is a clear evidence of age dispersion, with
clusters up to $\sim 25\%$ younger than the older members; and (c) the
clusters within the metal rich group ([Fe/H]$\geq-0.9$) seem to be coeval
within the uncertainties (except Pal~12), but younger ($\sim 17\%$) than
the bulk of the Galactic globulars. The latter result is totally model
dependent.

From the Galactocentric distribution of the GGC ages, we can divide
the GGCs in two groups: The old coeval clusters, and the young
clusters. The second group can be divided into two subgroups, the
``real young clusters'' and the ``young, but model dependent'', which
are within the intermediate and high metallicity groups,
respectively. From this distribution, we can present a possible
scenario for the Milky Way formation: The GC formation process started
at the same zero age throughout the halo, at least out to $\sim
20$~kpc from the Galactic center. According to the present stellar
evolution models, the metal-rich globulars are formed at a later time
($\sim 17\%$ lower age).  And finally, significantly younger halo GGCs
are found at any $R_{\rm GC}>8 kpc$. For these, a possible scenario
associated with mergers of dwarf galaxies to the Milky Way is
suggested.

\keywords{Hertzsprung-Russell (HR) Diagram -- Stars: Population II --
Globular Clusters: General -- The Galaxy: Evolution -- The Galaxy:
Formation.}

\end{abstract}

\section{Introduction}	
\label{intro}

Galactic globular clusters (GGC) are the oldest components of the Galactic
halo for which ages can be obtained. The determination of their relative
ages and of any age correlation with metallicities, abundance patterns,
positions and kinematics provides clues on the formation timescale of the
halo and gives information on the early efficiency of the enrichment
processes in the proto--Galactic material. The importance of these problems
and the difficulty in answering these questions is at the basis of the huge
efforts dedicated to gather the relative ages of GGCs in the last 30 years
or so [VandenBerg et al. (\cite{vsb96}), Sarajedini et al. (\cite{scd97}),
and references therein].

The methods at use for the age determination of GGCs are based on the
position of the turnoff (TO) in the color--magnitude diagram (CMD) of
their stellar population. We can measure either the absolute magnitude
or the de--reddened color of the TO. However, in order to overcome the
uncertainties intrinsic to any method to get GGC distances and
reddening, it is common to measure either the color or the magnitude
(or both) of the TO, relative to some other point in the CMD whose
position has a negligible dependence on age.

Observationally, as pointed out by Sarajedini \& Demarque
(\cite{sd90}) and VandenBerg et al. (\cite{vbs90}), the most precise
relative age indicator is based on the TO color relative to some fixed
point on the red giant branch (RGB). This  is usually called
the ``horizontal method''. Unfortunately, the theoretical RGB temperature
is very sensitive to the adopted mixing length parameter, whose
dependence on the metallicity is not well established, yet. As a
consequence, investigations on relative ages based on the horizontal
method might be of difficult interpretation, and need a careful
calibration of the relative TO color as a function of the relative age
(Buonanno et al. \cite{b98}).  
The other age indicator, the ``vertical method'', is based on the TO
luminosity relative to the horizontal branch (HB). Though this is usually
considered a more robust relative age indicator, it is affected both by the
uncertainty of the dependence of the HB luminosity on metallicity and the
empirical difficulties to get the TO, and the HB magnitudes for clusters
with only blue HBs.
It was also pointed out by Sweigart (\cite{sweigart1997}) and Sandquist et
al. (\cite{sand1999}), that there is the possibility that at a given
[Fe/H], there may be a dispersion in the content of helium in the envelope
HB stars in different clusters. 
At a given [Fe/H], this would lead to a
range in HB magnitude and add some scatter to the vertical method of relative age
determination.

It must also be noted that both methods are affected by the still
uncertain dependence of the alpha elements and helium content on the
metallicity.

Given these problems, it is still an open debate whether most GGCs are
almost coeval (Stetson et al. \cite{svb96}) or whether there was a
protracted formation epoch of 5 Gyr (Sarajedini et al. \cite{scd97}) or so
(i.e. for 30-40\% of the Galactic halo lifetime).

Indeed, there is a major limitation to the large scale GGC relative
age investigations: the photometric inhomogeneity and the
inhomogeneity in the analysis of the databases used in the various
studies. Many previous studies frequently combine photographic and CCD
data, different databases (obtained with different instruments with
uncertain calibrations to standard systems and/or based on different
sets of standards), or inappropriate color-magnitude diagrams (CMD)
were used. This inhomogeneity affects even many recent works, for
which results can not yet been considered conclusive (see Stetson et
al. 1996 for a discussion).

Recently, two new investigations have brought fresh views in this
field.  First, an analysis of published CMDs both in the $B,V$ and
$V,I$ bands was carried out in Saviane et al. (\cite{srp97}; hereafter
SRP97). SRP97 showed that the ($V-I$) TO-RGB color differences are
less sensitive to metallicity than the ($B-V$) ones (while retaining
the same age sensitivity). SRP97 also suggested that a high-precision,
large-scale investigation in the $V$ and $I$ bands would have allowed
a relative age determination through the horizontal method without the
usual limitation of dividing the clusters into different metallicity
groups (VandenBerg et al. 1990). Still, a calibration of the
horizontal methods in the $V$ and $I$ bands was needed for a correct
interpretation of the data.

Later on, Buonanno et al. (\cite{b98}) showed that, with an
appropriate calibration based on the vertical method, reliable
relative ages can indeed be obtained with the horizontal method. The
investigation of Buonanno et al. (\cite{b98}) is based both on
original and literature $(B-V)$ material.

The results presented here take advantage of the strengths of both
investigations. Soon after the SRP97 study, we began the collection of
an homogeneous photometric material for a large sample of GGCs, in
order to obtain accurate relative ages by using the horizontal method
in the $(\rm [Fe/H], \delta (V-I))$ plane. Our first observational
effort, aimed at the inner-intermediate halo clusters, is now
complete, and we provide here the first results.

In the next section, the data used for this study are presented. In
Sect.~\ref{method} we define our age indicators and explain how they
have have been measured on both the CMDs and theoretical
models. Section~\ref{relages} presents the measures obtained following
this procedure and compares them with the predictions of the
theoretical models.  In Sect.~\ref{discussion} we discuss our
results. An analysis of the relative ages versus the metallicity
(Sect.~\ref{MetalDist}) and Galactocentric distance
(Sect.~\ref{GalDist}) is presented. The discussion is also carried out
comparing clusters in metallicity subgroups
(Sect.~\ref{metgroups}). In Sect.~\ref{discussion_formation} the clues
obtained till now are used to gather some information on the Milky Way
formation and evolution. A summary is finally given in
Sect~\ref{summary}. The potentiality of our data base for testing the
theoretical calculations is also discussed in Appendix~\ref{test}.

\section{The data}	
\label{data}

The goal of our observational strategy was to obtain color differences
near the TO region with an uncertainty $\leq 0.01$mag, which allows a
$\leq 1$Gyr age resolution. As a first step, we used 1-m class
telescopes to build a large reference sample including all clusters
within $(m-M)_V=16$.  The 91cm ESO/Dutch Telescope (for the southern
sky GGCs) and the 1m 
Isaac Newton Group/Jacobus Kapteyn Telescope
 (for the northern sky GGCs) were then
used to cover 52 of the scheduled 69 clusters.  Of the total sample,
only 34 were suitable for this study. The remaining objects were
excluded due to several reasons: differential reddening, small number
of member stars, large background contamination, bad definition of the
RGB or HB.  One or two overlapping fields were covered for each
cluster, avoiding the cluster center, especially when it is crowded. From
2500 to 20000 stars per cluster were measured. The typical CMD extends from
the RGB tip to $\geq 3$~magnitudes below the TO.  The final selected sample
is listed in Tab.~\ref{measures}. Cluster names are given in col.~2. The
assumed [Fe/H], which covers almost the entire GGC metallicity range
($-2.1\leq {\rm [Fe/H]}\leq -0.7$), is given in col.~3. The [Fe/H] values
were taken (unless otherwise stated) from Rutledge et al. (\cite{rhs97})
(their Tab.~2, column 6).  Column 4 lists the Galactocentric distance
(from Harris \cite{harris}), which extends from 2 to $\sim 20$ kpc.  The
following columns report our measures, as discussed in Sect.~\ref{relages}.

In our attempt to be as homogeneous as possible, we have adopted the
metallicities listed in Rutledge et al. (\cite{rhs97}). Their values
were in fact obtained from a large and homogeneous work based on the
Ca~II triplet, and calibrated both over the Carretta \& Gratton
(\cite{cg97}) and the Zinn \& West (\cite{zw84}) scale.

In this paper we adopt the Carretta \& Gratton (\cite{cg97}) values,
as their metallicity scale was obtained from high resolution CCD
spectra of 24 GGCs (20 of them are in common with our sample),
analyzed in a self-consistent way. The main results presented in the
following sections would not change adopting the Zinn and West
(\cite{zw84}) scale.

A detailed description of the observation and reduction strategies are
given in Rosenberg et al. (1999a,b Papers I $\&$ II), where the CMDs
for the whole photometric sample are also presented. Here suffice it to
say that the data have been calibrated with the same set of standards,
and that the absolute zero-point uncertainties of our calibrations are
$\leq 0.02$ mag for each of the two bands. Moreover, 
three clusters have been observed with both the southern and northern
telescopes, thus providing a consistency check of the calibrations: the
zero points are consistent within the calibration errors, and, most
important, no color term is found between the two data sets.

Only two well known young clusters, Pal~1 (Rosenberg et al.
\cite{alf1}) and Pal~12 (Rosenberg et al. \cite{alf2}, Paper III), have
been observed in the {\it V,I} bands deep enough to allow the measurement
of their TOs. Since Pal~1 has no HB stars, Pal~12 remains the only cluster
that allows an extension of the present work to very young clusters: for
this reason, it has been included in our analysis,
even if its photometry is not strictly homogeneous (different
equipment has been used) with that of the other clusters, though the
photometric calibration has been done using the same set of standards
(Landolt \cite{landolt}), and at the same level of accuracy.

\begin{figure} 
\psfig{figure=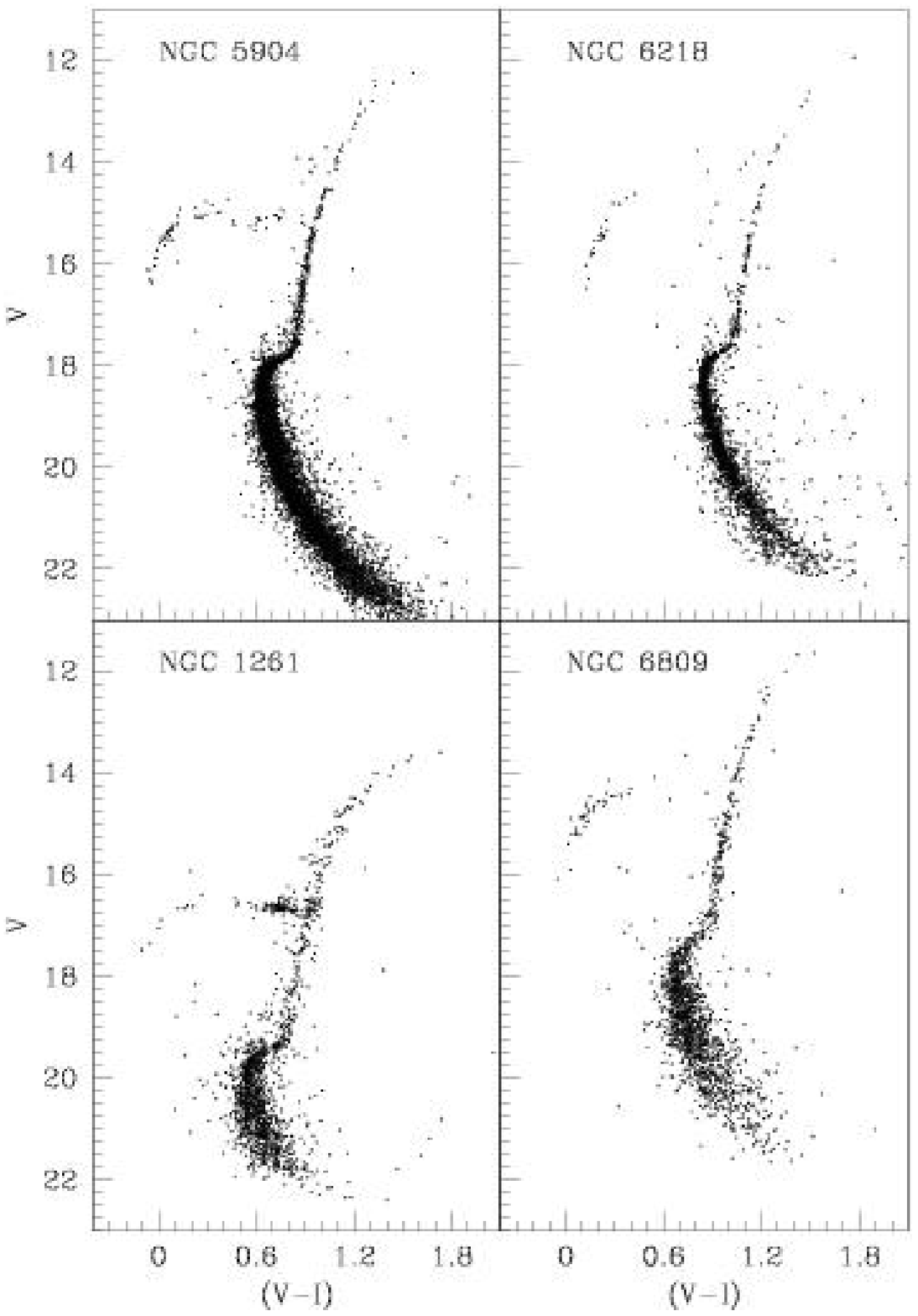,width=8.8cm}
\caption[]{The CMDs of 4 clusters used in the relative age determination,
which show the range in quality that is spanned by the present data
set. The top panels show two of the best CMDs (for the clusters NGC~5904
and NGC~6218); NGC~6809 and NGC~1261 (lower panels) are an example of lower
quality photometric samples.  In each case, the HB is populated by a good
number of stars, and more than 1 mag below the TO is covered even in the
case of NGC~1261.  }
\label{cmdquality}
\end{figure}

Figure~\ref{cmdquality} is an example of our photometry.  The CMDs of 4
clusters, representing diagrams covering the whole range in quality of our
data, are shown.

\section{Methodology}	
\label{method}

The key point ahead of the present analysis is the totally homogeneous
photometric sample that has been obtained.  There are several other
improvements with respect to previous investigations. In particular, (a) we
have used and analyzed three of the most recent evolutionary models;
(b) the theoretical trends of the photometric parameters have been
modeled with third-order polynomials in both the age and metallicity
instead of straight lines; and (c) a new and more homogeneous
metallicity scale (0.05~dex is the typical internal error on [Fe/H]),
calibrated on a large homogeneous spectroscopic sample, has been used.

We now discuss how the two observational databases were used to define
our differential age estimators, and how the theoretical models were
parameterized in order to convert our parameters into relative ages.

\subsection{Differential age estimators} 				%
\label{differential}

\begin{figure} 
\psfig{figure=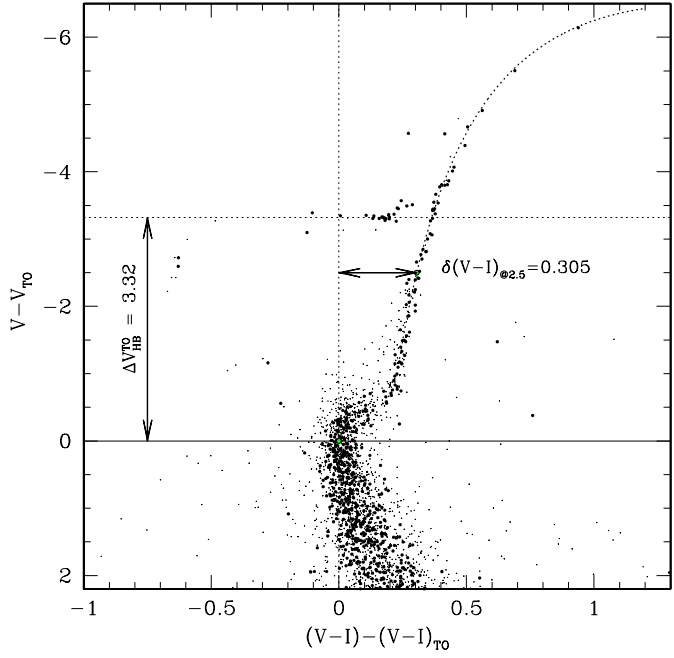,width=8.8cm}
\caption[]{
The CMD of NGC~1851. The heaviest points represent the selected CMD used to
measure the TO position and to fit the RGB fiducial line.  Magnitude and
color have been registered to the TO point. The vertical $\Delta V^{\rm
HB}_{\rm TO}$ and horizontal $\delta(V-I)_{@2.5}$ parameter values for this
cluster are indicated by arrows. The analytical fit to the RGB is also
shown.  }
\label{figdemo}
\end{figure}

Recent discussions on the possible choices for the photometric
parameters (which always measure the TO position with respect to some
other CMD feature with negligible dependence on age) can be found in
Stetson et al. (\cite{svb96}), Sarajedini et al. (\cite{scd97}) and
Buonanno et al. (\cite{b98}). Our investigation is based on two
``classical'' reddening and distance independent parameters: the
magnitude difference $\Delta V^{\rm HB}_{\rm TO}$ between the HB and
the TO (vertical method), and the color difference $\delta
(V-I)_{@2.5}$ between the TO and the RGB (horizontal method), where
the RGB color is measured 2.5 magnitudes above the TO. These
quantities are displayed in Fig.~\ref{figdemo} for NGC~1851.

A few other parameters, introduced in previous works, have been
measured and tested. VandenBerg et al. (1990) were the first to
suggest that the point on the MS 0.05 mag redder than the TO, could be
a better vertical reference point than the TO itself.  This point has
been consequently used for analyzing the magnitude difference relative
to the HB level (Buonanno et al. \cite{b98}) or as a reference point
for measuring the RGB-TO color difference 2.5 mag above it (VandenBerg
et al. \cite{vbs90}). We found this point useful for the very best
diagrams ($\sim10$ in our sample), but it is very difficult or
impossible to measure it for $\sim50\%$ of our clusters. Indeed, we
must recall that, from the observational point of view, we had to
reach a compromise between the deepness of our photometry and the size
of the sample that we could collect with a 1-m class telescope. As a
result, while the TO position can be reliably measured for all of our
selected clusters, the ``0.05'' point (which is $\sim1$~mag fainter
than the TO) generally falls in a MS region where the photometric
scatter is larger.

One might also question the $\Delta V=2.5$~mag choice and whether a
brighter point on the RGB could be better. To this respect, we must
consider that as we go from the TO up to the brighter part of the RGB,
the photometric error becomes smaller, but the RGB dependence on
[Fe/H] gets larger.  At the same time, the RGB is less and less
populated, so that it can be defined with a lower accuracy. In any
case, we made some tests by measuring the TO-RGB color difference for
magnitude offsets ranging from 1.5 to 3.5~mag above the TO.  We
concluded that the $\delta (V-I)_{@2.5}$ parameter represents the best
compromise.

\subsection{Measurement procedures}
\label{measurement}

In order to measure the morphological parameters, first the fiducial MS
lines were found by taking the median of the color distributions obtained
in magnitude boxes containing a fixed number of stars, ranging from 50 to
200 stars. The actual number was a function of the total number of stars
observed in the cluster.
This method allows to adapt the height of the magnitude box to the number
of stars that are found in each branch. It has for example the advantage
that the TO region, which has a strong curvature, can be sampled with a
small magnitude bin ($0.03 \div 0.04$~mag, tipically).

The RGBs were defined by fitting an analytic function to the fiducial
points starting from $\sim 1$~mag in $V$ above the TO. We found that a
hyperbolic function gives an excellent fit to these regions, being
able to follow the RGB trend even for the most metal rich clusters,
(Saviane et al. \cite{srpa99}). In particular, a function of the form:
\[
V = a + b \, (V-I) + c \, /  \, [(V-I) - d]
\]
was used. A dotted line shows the fit to the NGC~1851 RGB in
Fig.~\ref{figdemo}.

The HB level was found from the actual HB stars distribution for each
cluster by comparison with an empirically defined fiducial HB. The
latter was defined by starting with a bimodal HB cluster (NGC~1851),
and extending the HB to the red and to the blue by using our best
metal rich and metal poor clusters, respectively. Once the best fit
was found, the value $V_{\rm HB}$ was read at a color which
corresponds to $(V-I) = 0.2$ on the fiducial HB.

Finally, the turnoff position was found in a two-step
procedure. First, a preliminary location was defined by taking the
color and the magnitude of the bluest point on the fiducial MS lines;
then the color was fine-tuned by computing a statistics of the color
distribution near this point. All fiducial points whose colors are
within $\pm0.01$ mag of this preliminary TO position estimate were used to
compute the mean value which was assigned to the  TO. 
This step was iterated 20 times, keeping the color box fixed but changing
each time the stars that actually enter into the statistical computation,
according to the TO position. Usually, the procedure coverges very fast.

The measured values for the 35 GGCs are presented in
Tab.~\ref{measures}. The TO magnitudes and colors are given in cols.~5
and 6, while the obtained HB level is given in col.~7.

\subsection{Observational errors}

In order to estimate the uncertainty in the adopted TO color and
magnitude, we built a few hundred synthetic CMDs for each cluster,
using the Padova library of isochrones (see Bertelli et
al. \cite{b94}). These CMDs were done adopting for each cluster the
corresponding metallicity, the photometric errors (as estimated from
the star dispersion along the MS and lower SGB), and the total number
of stars in the observed CMD.  All synthetic models corresponding to a
given cluster were computed with the same input parameters, varying
only the initial random number generator seed.  The procedure used to
determine the TO (cf. Sec.~\ref{measurement}) was repeated for the
synthetic diagrams associated with each cluster, and the standard
deviation of the results was assumed to be the errors actually
affecting the color and magnitude of the TO in the observed CMDs.

The errors on the HB level are more difficult to estimate. As
explained before, the HB level was found using an empirically defined
fiducial HB. The usually small number of stars in this branch and
their non linear distribution with magnitude or color (from totally
red horizontal branches to nearly vertical blue HBs), does not allow
an easy estimate of the uncertainty associated with the HB magnitude.

The errors have been estimated by allowing the empirically defined
fiducial HB to move from the upper to the lower envelope of the HB in
each cluster. The uncertainties obtained in this way, 
turned out 
to be
similar among the clusters with red HBs and the clusters with blue
HBs, respectively.  Therefore, we decide to use a mean error of $\sim
0.05$ mag for the red HB objects, and $\sim 0.10$ mag for the blue
ones.  Note that these uncertainties must be considered an upper value
for the error, as among the stars in the brighter HB envelope there
are surely evolved HB stars.  Our HB level estimates are always within
0.1 mag of the Harris (\cite{harris}) compiled values, with the
exception of four clusters (NGC~6779, NGC~6681, NGC~6093, and
NGC~6254) for which more recent published photometry is found in
better agreement with our estimates than with Harris (\cite{harris}).

The estimated error for the RGB colors is the standard deviation of
the distribution of the residuals from the fiducial RGB of the color
of the stars located between 1.5 and 3.5 magnitudes above the TO. The
final error on $\Delta V^{\rm HB}_{\rm TO}$ is obtained as the
quadratic sum of the errors on the TO and HB magnitudes, while the
error on $\delta (V-I)_{@2.5}$ considers both the error in color and
magnitude of the TO (which affects the position of the reference point
on the RGB), and the error on the color of the point 2.5 mag brighter
than the TO magnitude.

\subsection{The theoretical models}					%
\label{teomod}

In order to interpret the results of our data samples, the theoretical
isochrones computed by Straniero et al. (\cite{scl97}, hereafter
SCL97), Cassisi et al. (\cite{cass98}, C98), and VandenBerg et
al. (\cite{vdb99}, V99) were used. These isochrones are the most
recent ones which provide $(V-I)$ colors and use updated physics. It
is important to notice that these theoretical models are completely
independent: indeed, they are obtained with different prescriptions
for the mixing-length parameter, the $Y$ vs. $Z$ relation, the
temperature-color transformations and bolometric corrections, etc.
The differences among the relative ages resulting from the models can
be taken as an indication of the (internal) uncertainties intrinsic to
our present knowledge of the stellar structure and evolution. The same
morphological parameters already defined for the observational CMDs
were measured on the isochrones.

The trends of the theoretical quantities as functions of both age and
metallicity were least-square interpolated by means of third-order
polynomials, so that the observed parameters can be easily mapped into
age and metallicity variations. The details of the fitting relations
are reported in Appendix~\ref{fitting}.

In order to calculate the theoretical values of $\Delta V^{\rm HB}_{\rm
TO}$, we have to assume a relation for the absolute $V$ magnitude of the HB
as a function of the metal content. In particular, here we adopted
$M_V(ZAHB) = 0.18 \cdot ({\rm [Fe/H]}+1.5) + 0.65$, from the recent
investigation of Carretta et al. (\cite{c99}). The implications of this
choice will be discussed in the following sections.

\section{Clusters' relative ages}					%
\label{relages}

In this section, relative ages are obtained from the observed $\Delta
V^{\rm HB}_{\rm TO}$ and $\delta (V-I)_{@2.5}$ parameters by comparison
with the V99 and SCL97 models. As discussed in the Introduction, from the
observational point of view, the horizontal method is a more precise
relative age indicator than the vertical one (Sarajedini et
al. \cite{sd90}, VandenBerg et al. \cite{vbs90}), as furtherly demonstrated
in Sec.~\ref{discussion}. Unfortunately, the dependence of the RGB
temperature on the the adopted mixing length parameter (whose dependence on
the metallicity is not well established yet), and the uncertain run of the
alpha elements enhancement, and helium content, with the metallicity (which
affect the vertical method as well) makes the data interpretation not
straightforward.  A detailed analysis of these effects is beyond the
purpose of the present paper.
However, we made an internal consistency check for the theoretical models,
and selected those for which the relative age trend with the metallicity
turned out
to be same (within the errors) using both the
horizontal and vertical method. While the V99 and SCL97 models satisfy
this condition (cf. Figs.~\ref{vertical_par} and \ref{horizontal_par})
for our sample of GGCs, C98 models do not.  Further tests are required
to identify the source of this problem, but it could be possibly
related to the $I$ bolometric corrections (cf. Appendix~\ref{test}),
so the C98 model predictions could still be valid for the $V$ and $B$
bands.  In any case, because of this internal inconsistency, from here
on we will base our analysis on the V99 and SCL97 models only. The
implications of the comparison between the observed data and the C98
models will be presented in Appendix~\ref{test}.

We want to note that the {\it absolute} ages obtained from the two
methods are not the same. The age differences between the vertical and
horizontal method are $\sim 1.2$ and $\sim 1.5$ Gyrs for the SCL and V99
models, respectively. This discrepancy can be removed by adopting for the
$V_{\rm HB}$ vs. [Fe/H] relation an appropriate constant. Far of being a
problem for our purpose of measuring relative ages, these discrepancies can
be a way to test the models and to fine-tune some still uncertain input
parameters. These points will be further discussed in Appendix~\ref{test}.
  
\subsection{Ages from the ``vertical'' method}				%
\label{Vvalues}

The measured $\Delta V^{\rm HB}_{\rm TO}$ parameter (and the
corresponding error) is listed for each cluster in Tab.~\ref{measures}
(col.~8). These values are plotted versus the cluster metallicity in
Fig.~\ref{vertical_par}. The dotted lines are the isochrones from the
V99 (top) and the SCL97 (bottom) models. Age is spaced by 1 Gyr steps,
with the lowermost line corresponding to 18 Gyrs.

We notice in the figure that the clusters are distributed in a narrow
band of $\le 2$ Gyrs width, apart from five clusters at [Fe/H] values
between -1.1 and -0.8 (namely, NGC~2808, NGC~362, NGC~1261, NGC~1851
and Pal~12). Within the observational errors, the theoretical
isochrones and the observed values show similar trends with
metallicity for $\rm [Fe/H] \leq -0.9$. It must be stated that this
result depends on the choice of the trend of the HB luminosity with
[Fe/H], although the conclusions would be the same if the slope of the
$V_{\rm HB}$ vs. [Fe/H] relation is changed by no more than $\pm 15\%$
(see also below). There is also a small second order dependence of the
relative ages on the zero-point of the relation, but this just changes
all the relative ages by a constant factor, while the trend with
[Fe/H] remains unchanged.

The isochrones were used to tentatively select a sample of coeval
clusters: first for each stellar evolution library the theoretical
locus that best fits the sample (not including Pal~12) was found, and
the relative $\Delta\,V_{\rm TO}^{\rm HB}$ with respect to this locus
were computed. We then chose to define as coeval GGCs those clusters
whose vertical parameter was within $\pm 1$ standard deviation from
the best-fitting isochrone. This interval is marked by thick lines in
Fig.~\ref{vertical_par}. Objects lying within this interval for both
sets of theoretical models (that we will call fiducial coeval from
here on) are marked by open circles in Fig.~\ref{vertical_par} and
will be used later on to test the isochrones in the $\delta
(V-I)_{@2.5}$ vs. [Fe/H] plane.  Interestingly enough, the same set of
coeval clusters is selected using both the SCL97 and the V99
isochrones, and using any slope $\alpha$ for the $V_{\rm HB}$
vs. [Fe/H] relation in the range $0.17<\alpha<0.23$ for the V99
isochrones and $0.15<\alpha<0.20$ for the SCL97 isochrones. The best
fitting isochrones have ages of 14.3 Gyrs according to the V99 models,
and 14.9 Gyrs from the SCL97 ones.
As it will be discussed below, the actual dispersion of the fiducial coeval
clusters around the mean isochrone, is indeed consistent with a null age
dispersion. 

We now turn our attention to those clusters that depart from the
distribution of the fiducial coevals. It must be noted that the
discrepancies are always in the sense of younger ages (smaller $\Delta
\, V_{\rm TO}^{\rm HB}$): moreover, for the discrepant clusters at
{\rm [Fe/H]}$\leq -0.9$, there are counterparts with similar
metallicity within the coeval sample, whereas for the more metal-rich
clusters the situation is less clear.  Indeed, if we rely on the
theoretical models, the three most metal rich clusters would seem
younger than 47~Tuc. However, it is well-known that problems arise in
modeling the RGB of metal rich stars (e.g. Stetson et
al.\cite{svb96}), so it could also be the case that the coeval cluster
band actually turns up at the metal rich end more than what is
predicted by the adopted models. We will have to come back to this
point later on.

For a better comparison of the results from the two methods, we
calculated what we call the mean normalized age. 
First, we derived the best mean age of the ``coeval'' clusters according to
each of the two sets of evolutionary models: viz., 14.3 Gyr for V99, and
14.9 Gyr for SCL97.  Then, for each cluster, we calculated the ratios of the
actual age for that cluster (as deduced from the model grids in the two
panels of Fig.~\ref{vertical_par}, cf. also Appendix A) relative to the mean
age, for the two cases.
The mean of these two normalized ages are listed in
Col.~3 of Tab.~\ref{ages}. The errors are the age intervals covered by
the photometric error bars in the normalized age scale.  In addition,
Col.~4 of the same Table gives the difference between the absolute
mean age of each cluster and the absolute age of the bulk of the GGCs,
assuming that the latter is 13.2 Gyrs as in Carretta et
al. \cite{c99}).

The age dispersions resulting from the vertical method are $\pm
1.4$~Gyr (independently from the adopted model) when using the entire
sample (excluding Pal~12); when only the fiducial coeval sample is
considered, the age dispersions become $\pm 0.7$ and $\pm 0.6$~Gyr
using SCL97 and V99 models, respectively. In terms of percent values,
this translates into a $9.2\%$ and $9.8\%$ (all clusters minus
Pal~12), and $4.4\%$ and $4.5\%$ (coeval sample) age dispersion. These
latter dispersions are fully compatible with the uncertainties in the
$\Delta V_{\rm HB}^{\rm TO}$ values, strengthening the idea that the
clusters selected as coeval must indeed have the same age.

\subsection{Ages from the ``horizontal'' method}			%
\label{Hvalues}

The measured $\delta (V-I)_{@2.5}$ parameters are presented in
Tab.~\ref{measures} (col.~9) and plotted versus the cluster
metallicity in Fig.~\ref{horizontal_par}. The dotted lines in the
figure represent the isochrones from V99 (upper panel) and SCL97
(lower panel), in 1~Gyr steps. The lowermost lines are the 18~Gyr and
17~Gyr isochrones, respectively.

Remarkably enough, Fig.~\ref{horizontal_par} resembles
Fig.~\ref{vertical_par}: again, most clusters are located in a narrow
sequence for ${\rm [Fe/H]}\leq -0.9$, with the exception of the same
previously identified five clusters, which result to have a younger
age also in this case. Also, the trend with metallicity is conserved,
with a similar uprise at the metal-rich end.

For the clusters at ${\rm [Fe/H]}\leq -0.9$~dex, the run of
$\delta(V-I)_{@2.5}$ is also reproduced by the isochrones. In this
metallicity range, the clusters selected as fiducial coeval by the
vertical method (open circles) still fall within a chronologically
narrow band of $\le 2$~Gyr, showing a remarkable consistency between
the two methods.

Apparently, the metal richer clusters are younger than the bulk of
Galactic globulars. 
Once more, this result is totally model-dependent,
and we must recall again that uncertainties in the color-temperature
relations and mixing length calibration,
as well as the run of the alpha elements content and helium abundance with
metallicity, 
 could affect the relative ages
obtained for the most metal rich objects (e.g. Stetson et
al. \cite{svb96}). Therefore, a problem with the theoretical relations
cannot be excluded, and NGC~104, NGC~6366, NGC~6352 and NGC~6838 could
indeed be coeval with the other clusters. Nevertheless, it must be
noted that the same trend is present on ages from the vertical method.
Moreover, if we apply a 0.07~mag correction for the HB magnitude of
the 4 most metal rich clusters (as suggested by Buonanno et al. 1998),
the ages obtained from the vertical method would be shifted towards
lower values, making them perfectly consistent with the results from
the horizontal method.  It is therefore tempting to consider the age
trend for the metal rich clusters to be a real possibility (which must
be furtherly tested with independent methods), although the precise
age offset remains to be established. In any case, if we take the
metal rich clusters as a single group, their internal age dispersion
is comparable to that of the rest of the fiducial coeval clusters.

As for the vertical method, normalized ages were obtained by means of
the difference in the $\delta (V-I)_{@2.5}$ parameter with respect to
the best fitting isochrones (13.1~Gyr and 16.4~Gyr for the SCL97 and
the V99 models, respectively). The resulting values are listed in
Tab.~\ref{ages} (cols.~5 and 6). In the table, the normalized ages
(col.~5) are the mean of the two values obtained using the two models,
while the age deviations in Gyr given in col.~6 are computed from
col.~5 assuming (as done in the previous section) a mean absolute age
of 13.2~Gyr (Carretta et al. \cite{c99}) for the mean age of the GGC
bulk. The errors are the age intervals covered by the photometric
error bars in the normalized age scale.

Since the $\delta(V-I)\div \delta t$ relation depends on the
metallicity in a non-linear way, the width covered by the $\pm 1$
standard deviation limits on the $\delta(V-I)_{@2.5}$ parameter (solid
lines on Fig.~\ref{horizontal_par}) is not constant. However, we find
that it goes from 0.010 to 0.007~mag, for the GGC metallicity range
$-2.1 \le {\rm [Fe/H]} \le -0.7$. This dispersion is comparable to the
experimental mean error for the coeval clusters (0.009~mag;
cf. Tab.~\ref{measures}).

Using the SCL97 models, the age dispersions that we have from the
horizontal method are $\sigma_t=1.2$~Gyr for the entire sample (with
the exception of Pal~12) and $\sigma_t=0.6$~Gyr (for the fiducial
coeval sample) corresponding to a percent age dispersion of $9.2\%$
and $4.3\%$. Similarly, from the data in the lower panel of
Fig.~\ref{horizontal_par} we have $\sigma_t=1.4$~Gyr and
$\sigma_t=0.6$~Gyr ($10.6\%$ and $4.5\%$). Although the absolute ages
of the clusters obtained from each model differ by $\sim 3$~Gyr, after
the normalization the relative ages are very similar. Moreover, these
relative ages are also close to those given by the vertical method
(cf. Section~\ref{Vvalues}).

As anticipated in the Introduction, Fig.~\ref{horizontal_par} shows a
mild metallicity dependence of the $\delta (V-I)$ parameter, smaller
than that of the corresponding $\delta (B-V)$ parameter (Buonanno et
al \cite{b98}). Indeed, as shown by Saviane et al. (\cite{srp97}),
and confirmed by Buonanno et al. (\cite{b98}), the slope of the
``isochrone'' in the $({\rm [Fe/H]}, \delta(B-V))$ plane is $\simeq
0.04$, while taking the coeval clusters at $\rm [Fe/H] < -1$ in
Fig.~\ref{horizontal_par} the slope of the isochrone is $\sim -0.025$.
This means that a typical error of 0.1~dex on the [Fe/H] translates in
a $\sim 0.4$~Gyr error on the relative cluster age if measured using
the traditional $(B-V)$ color, while it yields an error of 0.25~Gyr if
the age is measured with the present method.

Moreover, the self-consistency of the ages predicted by the two
methods and the two theoretical models strengthen the conclusions by
Saviane et al. (\cite{srp97}) that the $\delta (V-I)$ parameter is
much more reliable than the $\delta (B-V)$ as a relative age index. On
the contrary, using $\delta (B-V)$ and totally independent data sets,
both Saviane et al. (\cite{srp99}) and Buonanno et al. (\cite{b98})
show that significant discrepancies still exist between the ages
predicted by the vertical and horizontal methods.

As recalled in Sect.~\ref{intro}, the cluster Palomar~12 was included in
the present investigation, since it provides an excellent reference point
for the age calibration. It was found in Paper III that the age of this
cluster is $0.68 \pm 0.10$ that of both 47~Tuc and M5, as already suggested
by Gratton \& Ortolani (\cite{go88}) and Stetson et
al. (\cite{pal12stetson}). Here we find that the relative age of Pal~12
with respect to 47~Tuc is 0.68, while it is 0.62 with respect to M5, in
agreement with our previous investigation. This result is even more
striking if we take into
account that our old analysis was based on  three other independent
models.

\section{Discussion: Mean Age Distributions}	
\label{discussion}

\begin{figure} 
\psfig{figure=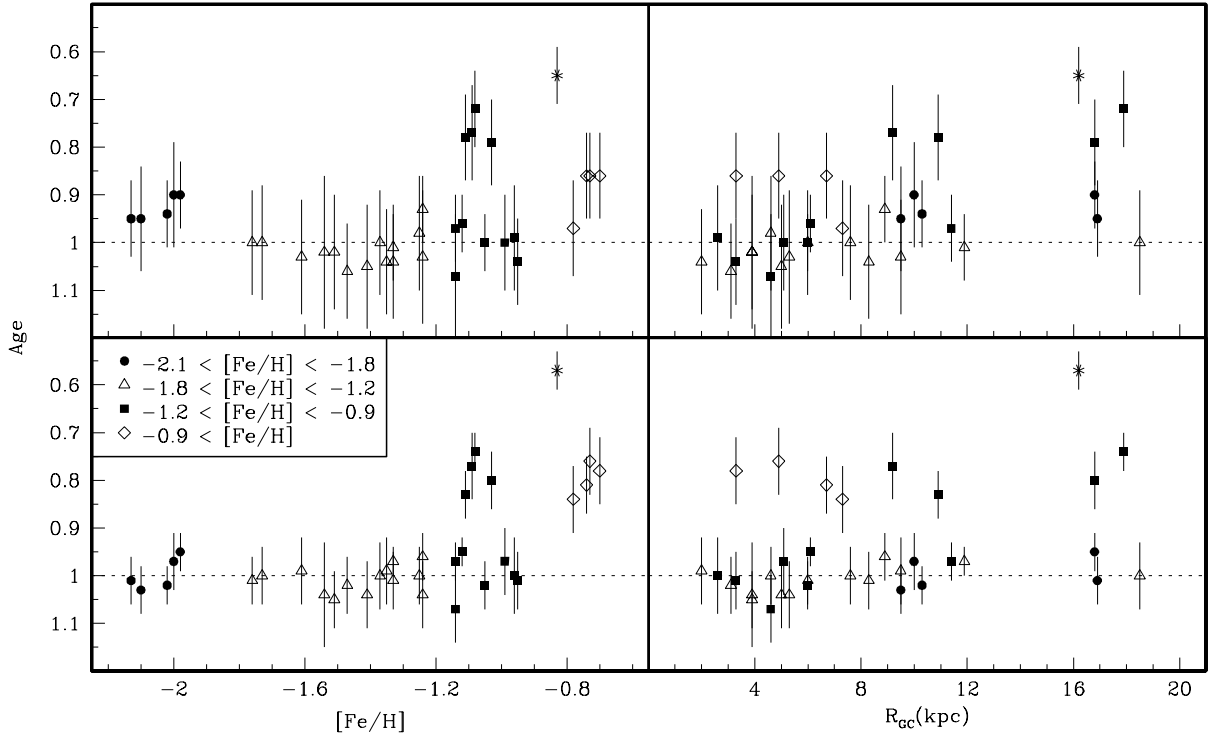,width=16cm}
\caption[]{
The normalized relative ages for our GGC sample from the vertical
(top panels) and the horizontal (bottom panels) methods are plotted
versus the metallicity ({\it left panels}) and versus the
Galactocentric distance ({\it right panels}). The different symbols
represent clusters in different metallicity groups as indicated in the
lower left panel. The error bars are the mean errors as given in
cols.~3 and 5 of Tab.~\ref{ages}. The youngest cluster (marked by an
asterisk) is Pal~12.}
\label{mean_age}
\end{figure}

\begin{figure} 
\psfig{figure=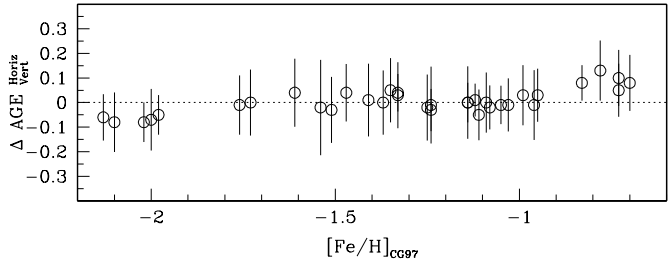,width=8.8cm}
\caption[]{The difference in the normalized relative ages obtained from the
two methods, $\Delta \rm Age_{Vert}^{Hor}$, as a function of the
metallicity. The error bars were obtained as the quadratic sum of the
errors from the two methods.
}
\label{cmp_ages}
\end{figure}

In this section, the age vs. metallicity and Galactocentric distance
trends will be discussed. We will use the normalized ages given in
Cols. 3 and 5 of Tab.~\ref{ages} for the vertical and horizontal
methods, respectively (the mean of these two values is given in
col.~7). Fig~\ref{mean_age} plots these normalized ages
vs. metallicity (left panels) and Galactocentric distance (right
panels). We arbitrarily divided our GGC sample into four metallicity
groups: (a) the very metal poor ($\rm [Fe/H]<-1.8$, filled circles),
(b) the metal poor ($-1.8 \leq [\rm Fe/H] < -1.2$, open triangles),
(c) the metal intermediate ($-1.2 \leq [\rm Fe/H] < -0.9$, filled
squares), and, finally, (d) the metal rich ($[\rm Fe/H]\geq -0.9$,
open diamonds). Notice that Pal~12 is always represented as an
asterisk.  The values from the vertical (upper panels) and horizontal
(lower panels) methods are plotted separately. Fig~\ref{mean_age}
shows two important features:
\begin{itemize}
\item 
The first one is that the general trend  shown by both
methods looks similar (within the errors).
A direct comparison of the two methods is provided in Fig.~\ref{cmp_ages},
where the difference in the normalized relative ages ($\Delta \rm
Age_{Vert}^{Hor}$) is plotted vs. the metallicity.  There is a very small
dispersion of $\Delta \rm Age_{Vert}^{Hor}$ around the zero level in each
metallicity group.  A marginal offset from the zero level for the two
extreme metallicity groups is present.  This could arise either from
discrepancies in the models and/or from the assumed relation for the
$V_{(\rm HB)}$ vs. [Fe/H] relation. If we were to act just on the HB level,
in order to have the same trend from the two methods, we should use a slope
$\Delta M_V{\rm (HB)}/\Delta \rm [Fe/H] = 0.08$ or 0.09 for the SCL97 and
V99 models, respectively. These values are not consistent with the current
estimates of this slope, so a partial correction of the (theoretical) TO
positions should also be considered. At this point, it is very important to
remark that the assumptions that must be introduced when using the vertical
method are not needed when working with the horizontal one. This method
relies on a minimum set of assumptions, thus making the interpretation of
the age rankings more straightforward. No parameterization of external
quantities (like the HB magnitude) is required.
\item 

The second important point is related to the observational errors.
The $\Delta V_{\rm HB}^{\rm TO}$ values are affected by uncertainties
that are $\sim 1.5-2.0$ times larger than those estimated for the
$\delta (V-I)_{@2.5}$ parameter. We already commented on the
possibility that our errors on $\Delta V_{\rm HB}^{\rm TO}$ could be
somehow overestimated (and this is also confirmed by the actual
dispersion of the points in Fig.~\ref{mean_age}).  On the other side,
though the observational errors on $\delta (V-I)_{@2.5}$ are surely
smaller, we still have to cope with the uncertainty (that we can not
estimate) on the theoretical colors when calculating the relative ages
with the horizontal method.  Still, as the observed trends from both
the vertical and horizontal methods are very similar, we prefer to
base our further discussion mainly on the results obtained from the
horizontal method, where the different trends and effects are more
clearly put into evidence.  In any case, it must be clearly stated
that the discussion would not change using the ages from the vertical
method.
\end{itemize}

\subsection{Distribution in Metallicity}				%
\label{MetalDist}

In Fig.~\ref{mean_age} (left panels) the fiducial normalized ages are
plotted vs. the cluster metallicities. Several regions of interest can
be discerned in the figure, and as a first step, we discuss here the
general trends that can be observed.

The dotted line represents the mean zero relative age level for the
coeval clusters: 26 out of 35 clusters are distributed around the mean
within an age interval $\Delta _{Age} \le 10\%$ of the mean. They all
have $\rm [Fe/H]<-0.9$.  In this region, no age-metallicity relation
is visible, when we take into account the errors on the ages. Within
the intermediate metallicity group, 4 clusters show definitely younger
ages than their equal metallicity counterparts, (namely; NGC~1261,
NGC~362 , NGC~2808 and NGC~1851). Notice also that no younger clusters
are detected for $\rm [Fe/H] < -1.2$.

The 5 clusters with the highest metallicities in our sample, have ages
significantly smaller than the mean age distribution. Of these, Pal~12
seems definitely younger than its equal metallicity counterparts. The
remaining 4 do not show any significant age dispersion. As already
discussed in Sect.~\ref{Hvalues}, this effect could be due to some
problems in the theoretical models at the metal rich end, but we must
note the internal consistency of the two methods. This can give some
support to the hypothesis that these 4 clusters might be really $\sim
17\%$ younger (and Pal~12 $\sim 40\%$ younger) than the bulk of GGCs.
These 4 objects are NGC~6366, NGC~6352, NGC~6838, and NGC~104. Notice
that assuming that these metal rich clusters are indeed younger would
have a strong consequence on the Galactic formation scenario, as we will
discuss in Section~\ref{discussion_formation}.

Taking the mean normalized ages (Tab.~\ref{ages}, col.7) within the
formerly defined metallicity groups, we find for the very metal poor
group a mean normalized age of $0.98\pm 0.03$, for the metal poor
group $1.01 \pm0.03$, for the metal intermediate $0.96 \pm 0.12$ if
the younger clusters are included and $1.00 \pm 0.04$ if they are not,
and for the metal rich $0.78 \pm 0.10$ if Pal~12 is included and $0.83
\pm 0.03$ if it is excluded. As it can be seen, the age dispersion
does not vary significantly along the metallicity range if only the
coeval clusters are considered. If one includes younger clusters into
the computation, then the metal intermediate group shows a larger age
dispersion. This is a well-known property of the GGCs (see
e.g. VandenBerg et al. \cite{vbs90}).

In conclusion, our data do not reveal an age-metallicity relation in
the usual sense of age decreasing (or increasing) with
metallicity. What is found is an increase of the age dispersion (due
to the presence of a few clusters with younger ages than the bulk of
the GGCs) for the metal rich clusters, while the lower metallicity
ones ([Fe/H]$\le -1.2$) seem to be all coeval.  This is in agreement
with the results of Richer et al. (\cite{r96}), Salaris \& Weiss
(\cite{sw98}), Buonanno et al. (\cite{b98}). On the other side,
Chaboyer et al. (\cite{cds96}) proposed an age-metallicity relation,
of the order $\Delta t_9/\Delta \rm [Fe/H] \simeq -4$~Gyr~dex$^{-1}$,
which is not present in our data set.  What happens for clusters with
[Fe/H]$\ge -0.9$ is totally model dependent; the models suggest a
younger age for these objects than for the more metal poor ones, and
no age dispersion.

\subsection{Testing young candidates within metallicity groups}
\label{metgroups}

Comparisons of relative ages have been often limited in the past to
clusters of similar metallicity. This indeed reduces the amount of
assumptions to be used, and allows an easier check of the relative
positions of the fiducial branches of the GGCs. Some ``template''
globular pairs or groups have many times been used in this
exercise. These special comparisons have been done mainly to establish
the efficiency of the halo formation but one could question whether
the detection of a single younger cluster can lead to any strong
conclusion in favor of some preferred Galactic halo formation
model. Aside from this consideration, we want to re-examine here some
of the special cases that have drawn much attention in the recent
past.

Our checks are made for metallicity groups. Again, the metallicity
scale is that of Carretta \& Gratton (\cite{cg97}): note that changing
the scale would change the absolute values of [Fe/H] but not the
membership to the metallicity groups. For each group, we will consider
those clusters that are significantly younger than other members of
the same group, or those objects which for any reasons have received
considerable attention in the recent past.

In order to make the reading of the next discussion easier, we will
abbreviate the principal papers in this way:  
Buonanno et al. 1998   		= B98, 
Chaboyer et al. 1996   		= C96,
Jonhson \& Bolte 1998  		= JB98,
Richer et al. 1996     		= R96, 
Salaris \& Weiss 1998  		= SW98, 
Sarajedini \& Demarque 1990 	= SD90,
Stetson et al. 1996    		= S96 and
VandenBerg et al. 1990 		= V90. 

\paragraph{
Very low metallicity group ($\rm [Fe/H] < -1.8$).}

For this group we found no evidence of age dispersion.  We will
comment on previous investigations (cf. B98, C96, SW98, V90 and R96)
for 4 globulars (NGC~4590, NGC~5053, NGC~6341 and NGC~7078). B98 and
R96 assign a younger age to NGC~4590 NGC~5053 and NGC~6341, and C96
agree that the former two should be younger. On the other side, SW98
and V90 find that the very metal poor clusters are all coeval within
the errors. Our Table~\ref{ages} formally indicate that NGC~4590 and
NGC~5053 are slightly younger than the other two; however these
differences are smaller than the quoted errors, and therefore not
significant.

\paragraph{Low metallicity group ($-1.8 \leq  {\rm [Fe/H]} < -1.2$).}

For this group we conclude that there is no evidence of age spread.
We consider the clusters NGC~5272 (M3), NGC~6205 (M13) and
NGC~1904. As in previous studies, we find that M13 results formally
older than M3, with NGC~1904 between the two, but these differences
are still within the observational errors, and therefore not
significant. On the contrary, C96 find M13 as much as $\sim 2$~Gyr
older than M3, but the recent accurate photometry of JB98 agrees with
our and earlier results.

\paragraph{
Intermediate metallicity group ($-1.2 \leq  {\rm [Fe/H]} < -0.9$).}

For this group we found clear evidence of age dispersion, with
clusters up to $\sim 25\%$ younger than the older members of the
group. We will center our attention on NGC~1851, NGC~1261, NGC~288,
NGC~2808 and NGC~362. As in the present paper, NGC~2808 is found to be
younger by previous investigations (C96, R96 and B98). NGC~1851 is
found young by C96, B98, SW98, R96 and the present work, while S96
claim that NGC~1851, NGC~362 and NGC~288 are coeval.

NGC~288 and NGC~362 have been often compared in the past: apart from C96,
all the previous investigations were based on the CMD obtained by Bolte
(\cite{bolte87}, \cite{bolte89}). Bolte (\cite{bolte89}), C96, R96, V90,
and SD90 claim that NGC~362 is significantly younger than NGC~288 (a $\sim
15-20\%$ lower age, in agreement with our result). A different
interpretation of the same data is offered by B98 and SW98, who did not
find significant age differences.  Still, most of the past studies agree
with our finding of a somewhat lower age for NGC~362 with respect to
NGC~288.
In the case of NGC~1261, apart from C96 (based on
Ferraro \cite{fer1261}), past investigations were based on the CMD
published by Bolte \& Marleau (\cite{bm89}). We find that this cluster
is $\sim 25\%$ younger than NGC~288, and this result goes in the same
sense of C96, R96 and Bolte (\cite{bolte89}), although the size of the
age offset is different. In contrast, B98 find no age difference and
SW98 find the cluster even older than NGC~288. It is difficult to
identify the origin of the difference with respect to the last two
investigations, since no value for the age indicators is given by
SW98, and B98 use $V_{05}$ as representative of the TO luminosity:
since the Bolte \& Marleau CMD becomes quite confused just below the
TO level, it is possible that the B98 value is affected by a large
error.  On the contrary, our CMD is better defined and more populated,
allowing a more reliable definition of the fiducial branches.  For
comparison, our $\Delta V^{0.05}$ estimate would be 0.25 mag brighter
than in B98, i.e. we would still find a younger age.

\paragraph{
High metallicity group (${\rm [Fe/H]} \geq -0.9$).}  

Except for the case of Pal~12, our conclusion for this group is that
these clusters are coeval, within the uncertainties, and possibly
younger than the lower metallicity ones. Most previous studies also
determined a constant age for this group, with the only exception of
C96. For NGC~104 and NGC~6838 all previous studies used the same
datasets (i.e. Hesser et al. \cite{h87} and Hodder et
al. \cite{hodd92} for the two clusters, respectively), while in the
case of NGC~6352 the Fullton et al. (\cite{full95}) CMD was used by
C96 and R96, and that of Buonanno et al. (\cite{b97}) was used by SW98
and B98. We can therefore take the C96 discrepant result as a sign of
the inherent uncertainties of the combined photometric databases and
measurement procedures. Indeed, the SW98, B98 and the present ages,
which are based on two independent methods, are all in fairly good
agreement.

\subsection{Radial Distribution of Age}				%
\label{GalDist}

Some important clues on the Milky Way formation and early evolution
can be obtained from the Galactocentric radial distribution of the
GGC relative ages. It is represented in Fig.~\ref{mean_age} (right
panels) and covers the Galactic zone between 2 and 18.5~kpc. The
$R_{\rm GC}$ values have been taken from Tab.~\ref{measures}.

We can clearly distinguish two groups of clusters: the old (coeval)
and a smaller sample of younger clusters. The two groups are better
seen in the lower panel (but see comments on the errors associated with
the vertical parameter in Section~\ref{Vvalues}).

We begin our discussion with those clusters significantly younger than
the bulk. They have at least a $10\%$~younger age. Within this group
we should distinguish between the ``really younger'' (NGC~1261,
NGC~1851, NGC~2808, NGC~362 and Pal~12), which have an older
counterpart at the same metallicity which turns out to be coeval with the
other most metal poor objects, and those lacking an old counterpart
with similar metallicity, for which the younger age is deduced by
comparison with the models, and hence is model dependent (NGC~104,
NGC~6352, NGC~6366 and NGC~6838). In the last (most metal rich) group,
four of the five clusters lie within 8~kpc from the Galactic center.
A young age for three of them was already suggested by Salaris \&
Weiss (\cite{sw98}), who find, as we do, an almost null age difference
within this group, and an average age $\sim 20\%$ younger than the
metal-poor halo clusters.

Beyond 8~kpc, five younger clusters are seen in Fig.~\ref{mean_age},
namely NGC~362, NGC~2808, Pal~12, NGC~1851 and NGC~1261 (in order of
increasing $R_{\rm GC}$).

Coming to the bulk of our cluster sample, we already noticed that for
the coeval clusters there is a small age dispersion around the mean
zero level ($\sim 4\%$ for the coeval sample), which is consistent
with a null dispersion when we take into account the observational
errors. This dispersion is much larger if we consider the whole
sample, but we do not find any Galactocentric distance vs. age
relation. However, it is interesting that, if the (uncertain) metal
rich clusters (marked by open diamonds in Fig.~\ref{mean_age}) were
excluded, it would appear that the age spread increases with the
Galactocentric radius.  This result has been reached also by Richer et
al. (\cite{r96}), Chaboyer et al. (\cite{cds96}), Salaris \& Weiss
(\cite{sw98}), Buonanno et al. (\cite{b98}), who include clusters out
to 100~kpc, 37~kpc, 27~kpc and 28~kpc respectively. All these studies
remark that younger clusters are present only in the outer regions.

In summary, as a matter of fact, the following picture arises from our
analysis:

\begin{itemize}

\item  According to the current models, most of the clusters are coeval 
and old.

\item A fraction of the intermediate metallicity and all the metal rich
clusters (according to the current models) are substantially younger.

\item The younger intermediate metallicity  clusters have all $R_{\rm GC}>8$~kpc.

\item The young clusters located at larger $R_{\rm GC}$ have typical halo
kinematics.

\end{itemize}

The consequences of these results on the mechanism of halo formation
are discussed in the next section.

\section{Clues on the Milky Way Formation} 
\label{discussion_formation}

\begin{figure} 
\psfig{figure=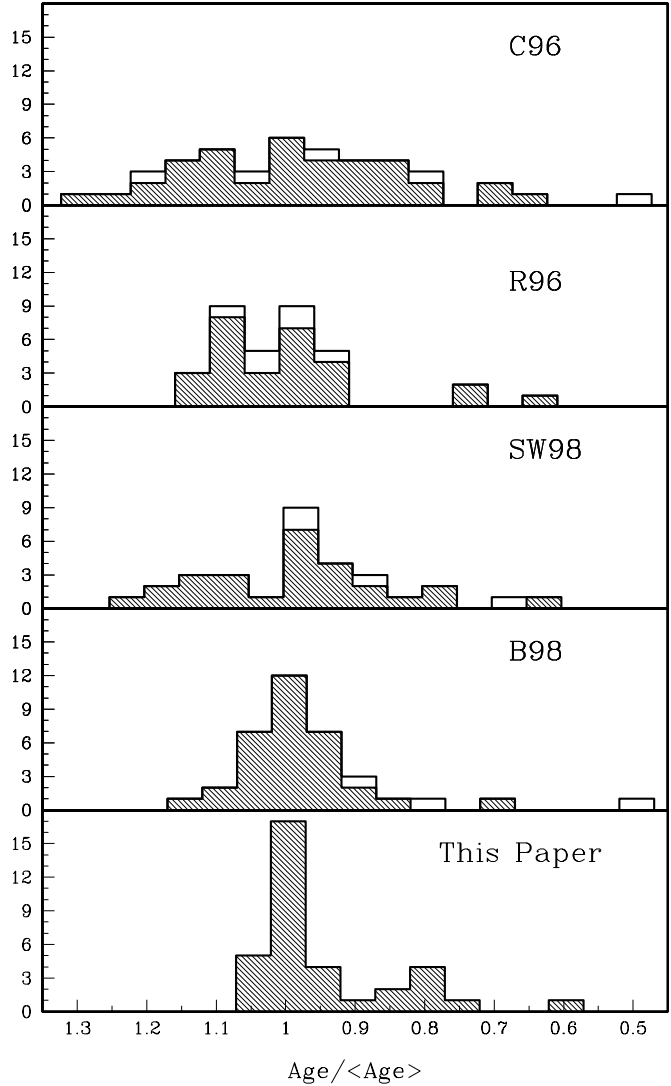,width=8.8cm}
\caption[]{ 
Histograms of the relative age distribution from the most recent
compilations in the literature. The histograms are centered on the
mean age of the respective samples.  Clusters located at the right are
younger. The shadowed zone represent the histogram for clusters within
20 kpc from the center of our Galaxy. The labels identify previous
investigations, as explained in the text. }
\label{histo_age}
\end{figure}

Fig.~\ref{histo_age} shows how the mean normalized relative ages
(Col.~7 of Table~\ref{ages}) compare with previous large-scale
investigations: the different panels show, from top to bottom,
histograms of the normalized age distributions found by Chaboyer et
al. (\cite{cds96}), Richer et al. (\cite{r96}), Salaris \& Weiss
(\cite{sw98}), Buonanno et al. (\cite{b98}), and the present study. In
order to intercompare them, they have been normalized to the mean
absolute age in each author's scale. For each histogram, the shadowed
area corresponds to GGCs with a Galactocentric distance smaller than
20~kpc.

It is clear that the age distributions become narrower as we go from older
to more recent studies. This is just the sign of the increasing accuracy of
the data samples, of the measurement procedures, and of the analysis
techniques. The principal improvements that we have introduced are: (a) the
use of the largest homogeneous CCD database (meaning with homogeneous that
the same instrumentation has been used, the same data and photometric
reduction procedures have been followed for all clusters, the same
calibration standards have been adopted, etc...); (b) the use of two
independent methods for the age measurement; (c) the use of $V$, $I$
photometry, and (d) a homogeneous metallicity scale and recent theoretical
models are also introduced.

The age dating progress that has been discussed so far has important
consequences on our interpretation of the timescales of the Milky Way
formation. In particular, we go from a halo formation lasting for
$\sim 40\%$ of the Galactic lifetime (C96), to the present result of
most of the halo clusters being coeval.

Besides this basic result, other clues on the Milky Way formation have
been obtained from the previous discussion. Going back to
Fig.~\ref{mean_age}, a chronological order of structure formation can
be inferred. The first objects to be formed are the halo clusters. Old
clusters are found at any distance from the Galactic center.

The GC formation process then started at the same zero age throughout
the halo, at least out to $\sim 20$~kpc from the center. All the more
metal rich ([Fe/H]$\geq-0.9$) clusters formed at later times ($\sim 17
\%$ of the halo age).  Once again, we stress that this interpretation
is model dependent, as it depends on the behavior of the isochrones at
high metallicities, and it is based on only 5 objects. Note that these
clusters do not identify a unique substructure of the Galaxy. One
(Pal~12), likely two (including NGC~6366, cf. Da Costa \&
Armandroff (\cite{da95})) are halo members, one might be a member of
the bulge population (NGC~6352, Minniti \cite{mi95}), and the last two
(NGC~6838 and 47~Tuc) of more uncertain classification, either thick disk
members (Armandroff \cite{a89}) or halo clusters crossing the disk,
following Minniti (\cite{mi95}) who showed that there is no thick disk
GGC population.

Finally, significantly younger halo GGCs are found at any $R_{\rm
GC}>8$~kpc.  These clusters (Pal~12, NGC~1851, NGC~1261, NGC~2808 and
NGC~362) could be associated with the so-called ``streams'', i.e. alignments
along great circles over the sky, which could arise from these
clusters being the relics of ancient Milky Way satellites of the size
of a dwarf galaxy (e.g. Lynden-Bell \& Lynden-Bell \cite{lyn95}, Fusi
Pecci et al. \cite{f95}).

\section{Conclusions}	
\label{summary}

Based on a new large, homogeneous photometric database for 34 Galactic
globular clusters (+ Pal~12), a set of distance and reddening
independent relative age indicators has been measured. The $\delta
(V-I)_{@2.5}$ and $\Delta V^{\rm HB}_{\rm TO}$ vs. metallicity
relations have been compared to the relations predicted by two recent
updated libraries of isochrones. Using these models and two
independent methods, we have found that self-consistent relative ages
can be estimated for our GGCs sample.  In turn, this demonstrates that
the two adopted models are internally self-consistent.

Based on the relative age vs. metallicity distribution, we conclude
that there is no evidence of an age spread for clusters with
[Fe/H]$<-1.2$,  all 19 clusters of our sample in this metallicity
range being old and coeval. For the intermediate metallicity group
($-1.2\leq $[Fe/H]$<-0.9$) there is a clear evidence of age
dispersion, with clusters up to $\sim 25\%$ younger than the older
members. Seven of the 11 GGCs in this group are coeval (also with the
previous group), while the remaining 4 are much younger (namely
NGC~362, NGC~1261, NGC~1851 and NGC~2808). Finally, the metal rich
group ([Fe/H]$\geq -0.9$) seem to be coeval within the uncertainties
(except Pal~12), and younger ($\sim 17\%$) than the rest of the
clusters,  this result being model dependent.

From the Galactocentric distribution of the GGC ages, we can divide
the GGCs in two groups, the old coeval clusters, and the young
clusters. The second group should be divided in two subgroups, the
``real young clusters'' and the ``model dependent'', located in the
intermediate and high metallicity groups, respectively. From this
distribution, we can present a possible interpretation of the Milky
Way formation:
\begin{itemize}

\item 
The GC formation process started at the same zero age throughout the
halo, at least out to $\sim 20$~kpc from the Galactic center.
\item 
At later ($\sim 17\%$ lower) times the metal-rich globulars are formed
(we stress that this interpretation is model dependent).
\item 
Finally, significantly younger halo GGCs are found at any $R_{\rm
GC}>8kpc$, for which a possible scenario associated with mergers of
dwarf galaxies to the Milky Way could be considered.
\end{itemize}
\appendix

\section{Theoretical model fitting}					%
\label{fitting}

As already introduced in Sect.~\ref{teomod}, and in order to interpret
the results of our data samples, the theoretical isochrones computed
by Straniero et al. (\cite{scl97}, SCL97), Cassisi et
al. (\cite{cass98}, C98), and VandenBerg et al. (\cite{vdb99}, V99)
were used. On these isochrones, the same morphological parameters
already defined for the observational CMDs ($\Delta V^{\rm HB}_{\rm
TO}$ and $\delta (V-I)_{@2.5}$), were measured.

The trends of the theoretical quantities as a function of both age and
metallicity were least-square interpolated by means of third-order
polynomials, so that the observed parameters can be easily mapped into
age and metallicity variations. This will allow us to easily translate
the parameter values into ages.

The equations used are of the form:
\begin{center}
$Parameter = a + b\cdot {\rm [Fe/H]} + c\cdot (log~t) + d\cdot {\rm
[Fe/H]}^2 + e\cdot (log~t)^2 + f\cdot {\rm [Fe/H]}\cdot (log~t) +
g\cdot {\rm [Fe/H]}^3 + h\cdot (log~t)^3 + i\cdot {\rm [Fe/H]}^2\cdot
(log~t) + j\cdot {\rm [Fe/H]}\cdot (log~t)^2$,
\end{center}
where $parameter$ represents one of the two photometric age indices
($\Delta V^{\rm HB}_{\rm TO}$ or $\delta (V-I)_{@2.5}$) and $t$ is the
age in Gyr. 

The [Fe/H] of the V99 models were provided by the authors, while for
the SCL97 and C98 models they were defined as $\rm [Fe/H] = \log
(Z/Z_\odot)$, setting $Z_\odot = 0.02$. The resulting coefficients are
listed in Tab.~\ref{coeff}, where the last line also reports the $rms$
of the fits in magnitude and age.

\begin{figure} 
\psfig{figure=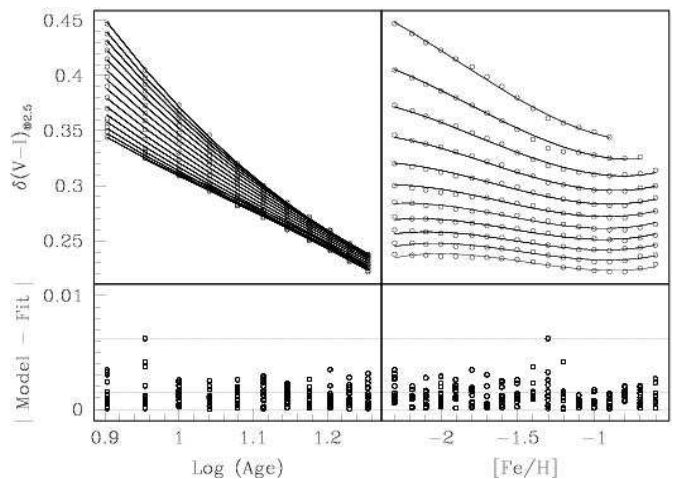,width=8.8cm}
\caption[]{
Example of our fit to the SCL97 models. The measured values on the
theoretical models are fitted in both the $\delta (V-I)_{@2.5}$ vs
[Fe/H] plane (upper-right panel) and the $\delta (V-I)_{@2.5}$
vs. $\log t$ plane (upper-left panel). The fit to the theoretical
values (open circles) are shown as continuous lines. The bottom panels
show the residuals}
\label{fitmodel}
\end{figure}

An example of our fits can be seen in Fig.~\ref{fitmodel}. The upper
left panel shows the $\delta (V-I)_{@2.5}$ vs. $\log$~age theoretical
behavior (at constant [Fe/H]), while the upper right panel shows the
same parameter vs. [Fe/H] (at constant ages). In both panels, model
results are shown by open circles, while our fits are represented by
solid lines. In the lower panels, the absolute residuals of the
respective fits are presented. The maximum difference between the
model and our fit is 0.006~mag, and the standard deviation $\sim
0.0015$~mag, which correspond to 0.15~Gyr. The dotted lines
graphically represent these two values.

A second order polynomial would not be able to follow the theoretical
trend of the models, while the distribution of the residuals shows
that a fourth order is not required, since the residual uncertainty is
much smaller than the observational error.

\section{A test bench for the theoretical models}			%
\label{test}

\begin{figure} 
\psfig{figure=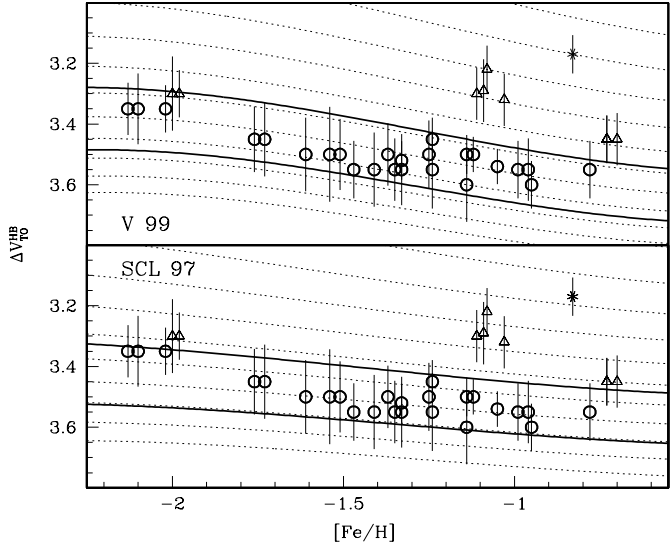,width=8.8cm}
\caption[]{ The measured $\Delta V^{\rm HB}_{\rm TO}$ parameter is
plotted
versus the metallicity. The dotted lines in the two panels show the
theoretical trend for V99 ({\it top}) and SCL97 ({\it bottom})
models. The isochrones are spaced by 1 Gyr (starting from 18 Gyr at
the bottom).  The asterisk represents the cluster Pal~12. The two
isochrones displayed as solid lines represent the $\pm 1$ standard
deviation limits of the $\Delta V^{HB}_{TO}$ parameter for the entire
sample (excluding Pal~12), and clusters falling within these
(circles), are defined as ``fiducial coeval''.  Note that the two
independent models give the same fiducial coeval object
selection.}
\label{vertical_par}
\end{figure}

\begin{figure} 
\psfig{figure=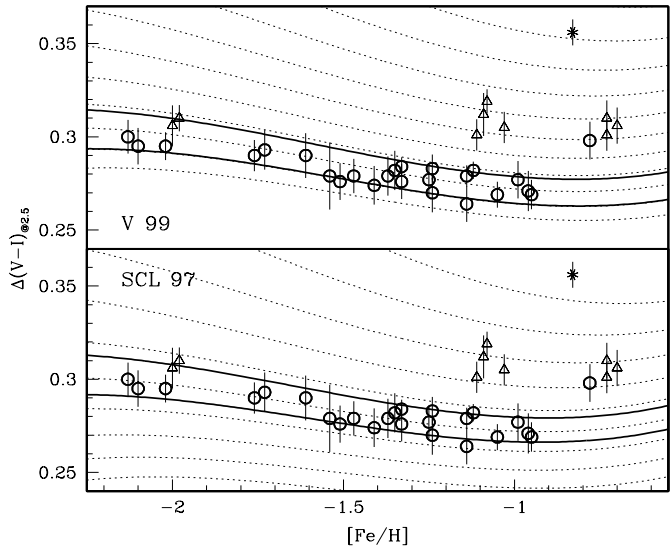,width=8.8cm}
\caption[]{ The measured $\delta (V-I)_{@2.5}$ parameter is plotted
versus metallicity. The same two sets of theoretical models of
Fig.~\ref{vertical_par} are shown (dotted lines). Age is spaced in 1
Gyr steps, the lowermost line corresponding to 18~Gyr and 17~Gyr
isochrones, for V99 and SCL97, respectively. The fiducial coeval
clusters selected in Fig.~\ref{vertical_par} are plotted as {\it open
circles}.  The two isochrones displayed as solid lines represent the
$\pm1$ standard deviation limits of the $\delta (V-I)_{@2.5}$
parameter for the GGC sample (except Pal~12).  Notice that the two
clusters at [Fe/H]$=-0.73$ (NGC~6366 and NGC~6838) have the same
$\Delta V_{\rm TO}^{HB}$, so they appear as a single point in
Fig.~\ref{vertical_par}. }
\label{horizontal_par}
\end{figure}

One can look at Fig.~\ref{vertical_par} and \ref{horizontal_par} as
empirical calibrations of the two differential parameters $\Delta
V_{\rm TO}^{HB}$ and $\delta (V-I)_{@2.5}$ as a function of [Fe/H].
Assuming that the two differential parameters are controlled just by
the age and the metallicity, when the theoretical loci are superposed
to these two diagrams, the same age-metallicity relations must be
obtained in the two cases.

We have shown that this is true for the the Straniero et
al. (\cite{scl97}, SCL97) and VandenBerg et al. (\cite{vdb99}, V99)
models, which indeed yield the same (shallow) age-metallicity relation
both using $\Delta V_{\rm TO}^{HB}$ and $\delta (V-I)_{@2.5}$.

\begin{figure} 
\psfig{figure=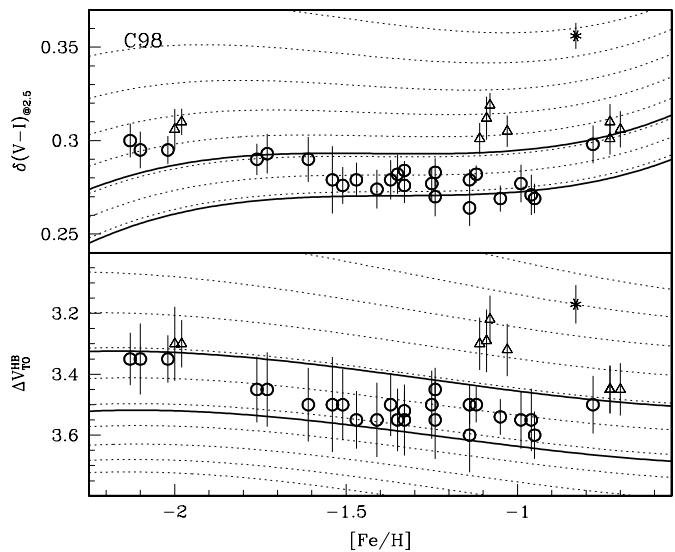,width=8.8cm}
\caption[]{
The same as in Figs.~\ref{vertical_par} and \ref{horizontal_par}, but
for
the models of C98.
}
\label{cas_mod}
\end{figure}

The same is not true for the C98 models: looking at Fig.~\ref{cas_mod}
it is clear that the theoretical isochrones show the same trend seen
for the other two sets of models in the lower panel (vertical method),
while, for example, an age-metallicity relation of $\sim +5$~Gyr/dex
for $\rm [Fe/H]<-1$ appears when the horizontal parameter is used
(inconsistent with the upper panel, and with what we have using SCL97
and V99 models). In order to reconcile the two diagrams, one could
play with the HB luminosity-metallicity relation. After a few tests,
we found that a partial agreement could be reached by using $M_V(\rm
HB)=0.35\,[Fe/H] + 1.40$, but such faint values for the RR~Lyr
luminosity are not consistent with the most recent results (see
e.g. Carretta et al. \cite{c99}), and the age-metallicity relation
would disagree in any case at the high [Fe/H] end.

We also checked the $B-V$ behavior of the horizontal parameter for
the C98 models, and in that case they agree with the SCL97 ones.

It is therefore suggested that the problems in
the C98 isochrones is related to the $I$ bolometric corrections (which
indeed are different from those used by both SCL97 and V99).

This test shows how our database can be used to define useful
observational constraints that any model calculation must
reproduce. Furthermore, we also suggest that a {\em multicolor}
approach should be followed to fully test the theoretical models.

\acknowledgments				

We thank Santino Cassisi and Alessandro Chieffi for providing us with
their models in tabular form. We are indebted to Don VandenBerg for
sending us his isochrones in advance of publication. We thank Vittorio
Castellani, Sergio Ortolani, Peter Stetson, and Don VandenBerg for the
useful discussions and encouragements. GP, IS, and AR acknowledge
partial support by the Ministero della Ricerca Scientifica e
Tecnologica and by the Agenzia Spaziale Italiana. AR has been
supported by the Italian Consorzio Nazionale Astronomia e Astrofisica.

\newpage

\begin{deluxetable}{crlrccccc}
\tablecolumns{9}
\tablewidth{0pt}
\tablenum{1}
\tablecaption{
Data for the 34 (+Pal~12) analyzed GGCs.
\label{measures}}
\tablehead{
\multicolumn{1}{c}{}                 & 
\multicolumn{1}{c}{Cluster}                      & 
\multicolumn{1}{c}{[Fe/H]}                       & 
\multicolumn{1}{c}{$R_{\rm GC}$}                 & 
\multicolumn{1}{c}{${V \rm (TO)}$}             &
\multicolumn{1}{c}{$\rm (V-I)_{(TO)}$}           & 
\multicolumn{1}{c}{${V \rm (HB)}$}             &
\multicolumn{1}{c}{$\Delta V_{\rm TO}^{\rm HB}$} &
\multicolumn{1}{c}{$\delta (V-I)_{@2.5}$} }
\tiny
\startdata
01 & NGC~104 & $-0.78\pm0.02 $ & 7.3 & $17.60\pm0.08$ &
$0.660\pm0.007$ & $14.05\pm0.05$ & $3.55\pm0.09$ & $0.295\pm0.010$ \nl
02 & NGC~288 & $-1.14\pm0.03 $ & 11.4 & $18.90\pm0.04$ &
$0.645\pm0.002$ & $15.40\pm0.05$ & $3.55\pm0.06$ & $0.276\pm0.006$ \nl
03 & NGC~362 & $-1.09\pm0.03 $ & 9.2 & $00.00\pm0.09$ &
$0.000\pm0.008$ & $03.29\pm0.05$ & $3.29\pm0.10$ & $0.312\pm0.011$ \nl
04 & NGC~1261 & $-1.08\pm0.04 $ & 17.9 & $19.90\pm0.06$ &
$0.555\pm0.003$ & $16.68\pm0.05$ & $3.22\pm0.08$ & $0.319\pm0.007$ \nl
05 & NGC~1851 & $-1.03\pm0.06 $ & 16.8 & $19.50\pm0.07$ &
$0.630\pm0.005$ & $16.18\pm0.05$ & $3.32\pm0.09$ & $0.305\pm0.008$ \nl
\hline 
06 & NGC~1904 & $-1.37\pm0.05  $ & 18.5 & $19.65\pm0.09$ &
$0.610\pm0.007$ & 
$16.15\pm0.05$ & $3.50\pm0.10$ & $0.279\pm0.010$ \nl 
07 & NGC~2808 & $-1.11\pm0.03  $ & 10.9 & $19.60\pm0.07$ &
$0.800\pm0.005$ & 
$16.30\pm0.05$ & $3.30\pm0.09$ & $0.301\pm0.008$ \nl
08 & NGC~3201 & $-1.24\pm0.03  $ &  8.9 & $18.20\pm0.05$ &
$0.905\pm0.004$ & 
$14.75\pm0.05$ & $3.45\pm0.07$ & $0.283\pm0.008$ \nl
09 & NGC~4590 & $-2.00\pm0.03  $ & 10.0 & $19.05\pm0.07$ &
$0.605\pm0.006$ & 
$15.75\pm0.10$ & $3.30\pm0.12$ & $0.306\pm0.010$ \nl 
10 & NGC~5053 & $-1.98\pm0.09  $ & 16.8 & $20.00\pm0.06$ &
$0.545\pm0.004$ & 
$16.70\pm0.05$ & $3.30\pm0.08$ & $0.310\pm0.007$ \nl
\hline 
11 & NGC~5272 & $-1.33\pm0.02^a$ & 11.9 & $19.10\pm0.04$ &
$0.575\pm0.002$ & 
$15.58\pm0.05$ & $3.52\pm0.06$ & $0.284\pm0.005$ \nl 
12 & NGC~5466 & $-2.13\pm0.36^b$ & 16.9 & $19.95\pm0.07$ &
$0.555\pm0.006$ & 
$16.60\pm0.05$ & $3.35\pm0.09$ & $0.300\pm0.009$ \nl 
13 & NGC~5897 & $-1.73\pm0.07  $ &  7.6 & $19.75\pm0.07$ &
$0.720\pm0.006$ & 
$16.30\pm0.10$ & $3.45\pm0.12$ & $0.293\pm0.011$ \nl
14 & NGC~5904 & $-1.12\pm0.03  $ &  6.1 & $18.50\pm0.03$ &
$0.625\pm0.002$ & 
$15.00\pm0.05$ & $3.50\pm0.06$ & $0.282\pm0.005$ \nl
15 & NGC~6093 & $-1.47\pm0.04  $ &  3.1 & $19.80\pm0.08$ &
$0.815\pm0.005$ & 
$16.25\pm0.05$ & $3.55\pm0.09$ & $0.279\pm0.009$ \nl 
\hline 
16 & NGC~6121 & $-1.05\pm0.03  $ &  6.0 & $16.90\pm0.03$ &
$1.125\pm0.004$ & 
$13.36\pm0.05$ & $3.54\pm0.06$ & $0.269\pm0.007$ \nl 
17 & NGC~6171 & $-0.95\pm0.04  $ &  3.3 & $19.25\pm0.06$ &
$1.150\pm0.004$ & 
$15.65\pm0.05$ & $3.60\pm0.08$ & $0.269\pm0.007$ \nl 
18 & NGC~6205 & $-1.33\pm0.05  $ &  8.3 & $18.50\pm0.06$ &
$0.575\pm0.004$ & 
$14.95\pm0.10$ & $3.55\pm0.12$ & $0.276\pm0.009$ \nl
19 & NGC~6218 & $-1.14\pm0.05  $ &  4.6 & $18.30\pm0.07$ &
$0.850\pm0.004$ & 
$14.70\pm0.10$ & $3.60\pm0.12$ & $0.264\pm0.010$ \nl 
20 & NGC~6254 & $-1.25\pm0.03  $ &  4.6 & $18.55\pm0.05$ &
$0.930\pm0.003$ & 
$15.05\pm0.10$ & $3.50\pm0.11$ & $0.277\pm0.009$ \nl 
\hline 
21 & NGC~6341 & $-2.10\pm0.02^a$ &  9.5 & $18.55\pm0.06$ &
$0.555\pm0.005$ & 
$15.20\pm0.10$ & $3.35\pm0.12$ & $0.295\pm0.010$ \nl 
22 & NGC~6352 & $-0.70\pm0.02  $ &  3.3 & $18.70\pm0.07$ &
$0.985\pm0.007$ & 
$15.25\pm0.05$ & $3.45\pm0.09$ & $0.306\pm0.010$ \nl
23 & NGC~6362 & $-0.99\pm0.03  $ &  5.1 & $18.90\pm0.08$ &
$0.685\pm0.007$ & 
$15.35\pm0.05$ & $3.55\pm0.09$ & $0.277\pm0.010$ \nl
24 & NGC~6366 & $-0.73\pm0.05  $ &  4.9 & $19.10\pm0.06$ &
$1.570\pm0.005$ & 
$15.65\pm0.05$ & $3.45\pm0.08$ & $0.310\pm0.009$ \nl 
25 & NGC~6397 & $-1.76\pm0.03  $ &  6.0 & $16.40\pm0.04$ &
$0.775\pm0.002$ & 
$12.95\pm0.10$ & $3.45\pm0.11$ & $0.290\pm0.008$ \nl
\hline 
26 & NGC~6535 & $-1.51\pm0.10  $ &  3.9 & $19.30\pm0.06$ &
$1.105\pm0.004$ & 
$15.80\pm0.10$ & $3.50\pm0.12$ & $0.270\pm0.010$ \nl
27 & NGC~6656 & $-1.41\pm0.03^a$ &  5.0 & $17.80\pm0.07$ &
$0.960\pm0.005$ & 
$14.25\pm0.10$ & $3.55\pm0.12$ & $0.274\pm0.010$ \nl 
28 & NGC~6681 & $-1.35\pm0.03  $ &  2.0 & $19.25\pm0.09$ &
$0.690\pm0.007$ & 
$15.70\pm0.05$ & $3.55\pm0.10$ & $0.282\pm0.011$ \nl
29 & NGC~6723 & $-0.96\pm0.04  $ &  2.6 & $19.00\pm0.09$ &
$0.725\pm0.007$ & 
$15.45\pm0.05$ & $3.55\pm0.10$ & $0.271\pm0.011$ \nl 
30 & NGC~6752 & $-1.24\pm0.03  $ &  5.3 & $17.35\pm0.08$ &
$0.705\pm0.005$ & 
$13.80\pm0.10$ & $3.55\pm0.13$ & $0.270\pm0.010$ \nl
\hline 
31 & NGC~6779 & $-1.61\pm0.13^b$ &  9.5 & $19.80\pm0.11$ &
$0.840\pm0.008$ & 
$16.30\pm0.05$ & $3.50\pm0.12$ & $0.290\pm0.012$ \nl
32 & NGC~6809 & $-1.54\pm0.03  $ &  3.9 & $17.95\pm0.12$ &
$0.680\pm0.014$ & 
$14.45\pm0.10$ & $3.50\pm0.16$ & $0.279\pm0.018$ \nl 
33 & NGC~6838 & $-0.73\pm0.03  $ &  6.7 & $17.95\pm0.06$ &
$0.935\pm0.005$ & 
$14.50\pm0.05$ & $3.45\pm0.08$ & $0.301\pm0.008$ \nl 
34 & NGC~7078 & $-2.02\pm0.04  $ & 10.3 & $19.25\pm0.06$ &
$0.650\pm0.004$ & 
$15.90\pm0.05$ & $3.35\pm0.08$ & $0.295\pm0.007$ \nl
35 & Pal 12   & $-0.83\pm0.06  $ & 16.2 & $20.35\pm0.06$ &
$0.695\pm0.005$ & 
$17.18\pm0.02$ & $3.17\pm0.06$ & $0.356\pm0.007$ \nl
\hline
\enddata
In the following cases, the [Fe/H] values were taken from: $(^a)$ CG97
and $(^b)$ ZW84 (transformed to the CG97 scale, as given by CG97).
\end{deluxetable}

\newpage

\begin{deluxetable}{cc|rr|rr|rr|c}
\tablecolumns{9}
\tablewidth{0.0pt}
\tablecaption{
Galctic globular clusters relative ages.
The last column indicates if the cluster is coeval (C), younger
(Y) or probably younger(Y?).
\label{ages}}
\tablenum{2}
\tablehead{ 
\multicolumn{2}{c}{Cluster name} 		& 
\multicolumn{2}{c}{Vertical Method} 		&
\multicolumn{2}{c}{Horizontal Method} 		& 
\multicolumn{2}{c}{Mean Age} 			&
\multicolumn{1}{c}{}                 		\\
\multicolumn{1}{c}{NGC} 			&
\multicolumn{1}{c}{Other} 			&
\multicolumn{1}{c}{$\overline {Age}$} 	& 
\multicolumn{1}{c}{$\Delta Age (Gyr)$}		&
\multicolumn{1}{c}{$\overline {Age}$} 	& 
\multicolumn{1}{c}{$\Delta Age (Gyr)$} 		& 
\multicolumn{1}{c}{$\overline {Age}$} 	&
\multicolumn{1}{c}{$\Delta Age (Gyr)$} 		& 
\multicolumn{1}{c}{}
}
\tiny
\startdata
104  & 47~Tucanae & $0.97\pm0.10$ & $-0.4\pm1.4$ & $0.84\pm0.07$ &
$-2.0\pm0.9$ 
& $0.90\pm0.08$ & $-1.2\pm1.2$ & Y? \nl
288  &  -         & $0.97\pm0.07$ & $-0.3\pm0.9$ & $0.97\pm0.04$ &
$-0.3\pm0.5$ 
& $0.97\pm0.05$ & $-0.3\pm0.7$ & C  \nl
362  &  -         & $0.77\pm0.10$ & $-2.9\pm1.4$ & $0.77\pm0.07$ &
$-2.9\pm1.0$ 
& $0.77\pm0.08$ & $-2.9\pm1.2$ & Y  \nl
1261 &  -         & $0.72\pm0.08$ & $-3.6\pm1.0$ & $0.74\pm0.04$ &
$-3.3\pm0.5$ 
& $0.73\pm0.05$ & $-3.5\pm0.8$ & Y  \nl
1851 &  -         & $0.79\pm0.09$ & $-2.7\pm1.2$ & $0.80\pm0.06$ &
$-2.5\pm0.7$ 
& $0.80\pm0.07$ & $-2.6\pm0.9$ & Y  \nl 
\hline 															   
1904 &  M~79      & $1.00\pm0.11$ & $ 0.0\pm1.4$ & $1.00\pm0.07$ & $
0.0\pm0.9$ 
& $1.00\pm0.08$ & $ 0.0\pm1.2$ & C  \nl 
2808 &  -         & $0.78\pm0.09$ & $-2.8\pm1.2$ & $0.83\pm0.05$ &
$-2.1\pm0.7$ 
& $0.81\pm0.07$ & $-2.5\pm0.9$ & Y  \nl 
3201 &  -         & $0.93\pm0.07$ & $-0.8\pm1.0$ & $0.96\pm0.05$ &
$-0.4\pm0.7$ 
& $0.95\pm0.06$ & $-0.6\pm0.8$ & C  \nl 
4590 &  M~68      & $0.90\pm0.11$ & $-1.2\pm1.5$ & $0.97\pm0.06$ &
$-0.3\pm0.7$ 
& $0.94\pm0.08$ & $-0.8\pm1.1$ & C  \nl 
5053 &  -         & $0.90\pm0.07$ & $-1.2\pm1.0$ & $0.95\pm0.04$ &
$-0.6\pm0.5$ 
& $0.93\pm0.05$ & $-0.9\pm0.7$ & C  \nl
\hline 															   
5272 &  M~3       & $1.01\pm0.07$ & $ 0.1\pm0.9$ & $0.97\pm0.03$ &
$-0.3\pm0.4$ 
& $0.99\pm0.05$ & $ 0.0\pm0.7$ & C  \nl 
5466 &  -         & $0.95\pm0.08$ & $-0.6\pm1.1$ & $1.01\pm0.05$ & $
0.1\pm0.6$ 
& $0.98\pm0.06$ & $-0.2\pm0.8$ & C  \nl 
5897 &  -         & $1.00\pm0.12$ & $ 0.0\pm1.6$ & $1.00\pm0.06$ & $
0.0\pm0.8$ 
& $1.00\pm0.09$ & $ 0.0\pm1.2$ & C  \nl
5904 &  M~5       & $0.96\pm0.06$ & $-0.4\pm0.8$ & $0.95\pm0.03$ &
$-0.6\pm0.4$ 
& $0.96\pm0.04$ & $-0.5\pm0.6$ & C  \nl
6093 &  M~80      & $1.06\pm0.10$ & $ 0.8\pm1.3$ & $1.02\pm0.06$ & $
0.3\pm0.8$ 
& $1.04\pm0.07$ & $ 0.5\pm1.0$ & C  \nl 
\hline 															   
6121 &  M~4       & $1.01\pm0.06$ & $ 0.0\pm0.8$ & $1.02\pm0.05$ & $
0.3\pm0.7$ 
& $1.01\pm0.05$ & $ 0.1\pm0.7$ & C  \nl 
6171 &  M~107     & $1.04\pm0.09$ & $ 0.5\pm1.1$ & $1.01\pm0.06$ & $
0.1\pm0.8$ 
& $1.02\pm0.07$ & $ 0.3\pm0.9$ & C  \nl 
6205 &  M~13      & $1.04\pm0.12$ & $ 0.5\pm1.6$ & $1.01\pm0.06$ & $
0.1\pm0.8$ 
& $1.02\pm0.09$ & $ 0.3\pm1.2$ & C  \nl
6218 &  M~12      & $1.07\pm0.13$ & $ 0.9\pm1.7$ & $1.07\pm0.07$ & $
0.9\pm0.9$ 
& $1.07\pm0.10$ & $ 0.9\pm1.3$ & C  \nl 
6254 &  M~10      & $0.98\pm0.12$ & $-0.2\pm1.5$ & $1.00\pm0.06$ & $
0.0\pm0.8$ 
& $0.99\pm0.08$ & $ 0.0\pm1.2$ & C  \nl 
\hline 															   
6341 &  M~92      & $0.95\pm0.11$ & $-0.6\pm1.4$ & $1.03\pm0.05$ & $
0.4\pm0.7$ 
& $0.99\pm0.08$ & $ 0.0\pm1.1$ & C  \nl 
6352 &  -         & $0.86\pm0.09$ & $-1.7\pm1.2$ & $0.78\pm0.07$ &
$-2.8\pm0.9$ 
& $0.82\pm0.08$ & $-2.3\pm1.1$ & Y? \nl
6362 &  -         & $1.00\pm0.10$ & $ 0.0\pm1.4$ & $0.97\pm0.07$ &
$-0.3\pm0.9$ 
& $0.99\pm0.08$ & $-0.1\pm1.1$ & C  \nl
6366 &  -         & $0.86\pm0.09$ & $-1.7\pm1.1$ & $0.76\pm0.07$ &
$-3.1\pm0.9$ 
& $0.81\pm0.07$ & $-2.4\pm1.0$ & Y? \nl 
6397 &  -         & $1.00\pm0.11$ & $ 0.0\pm1.4$ & $1.01\pm0.05$ & $
0.1\pm0.6$ 
& $1.00\pm0.07$ & $ 0.1\pm1.0$ & C  \nl
\hline 															   
6535 &  -         & $1.02\pm0.12$ & $ 0.3\pm1.6$ & $1.05\pm0.06$ & $
0.7\pm0.8$ 
& $1.03\pm0.09$ & $ 0.5\pm1.2$ & C  \nl
6656 &  M~22      & $1.05\pm0.13$ & $ 0.7\pm1.7$ & $1.04\pm0.07$ & $
0.5\pm0.9$ 
& $1.04\pm0.09$ & $ 0.6\pm1.3$ & C  \nl 
6681 &  M~70      & $1.04\pm0.11$ & $ 0.5\pm1.4$ & $0.99\pm0.07$ & $
0.0\pm0.9$ 
& $1.01\pm0.08$ & $ 0.2\pm1.2$ & C  \nl
6723 &  -         & $0.99\pm0.11$ & $ 0.0\pm1.5$ & $1.00\pm0.09$ & $
0.0\pm1.0$ 
& $1.00\pm0.09$ & $ 0.0\pm1.3$ & C  \nl 
6752 &  -         & $1.03\pm0.14$ & $ 0.4\pm1.8$ & $1.04\pm0.07$ & $
0.5\pm1.0$ 
& $1.03\pm0.10$ & $ 0.5\pm1.4$ & C  \nl
\hline 															   
6779 &  M~56      & $1.03\pm0.12$ & $ 0.4\pm1.6$ & $0.99\pm0.07$ & $
0.0\pm1.0$ 
& $1.01\pm0.09$ & $ 0.1\pm1.3$ & C  \nl
6809 &  M~55      & $1.02\pm0.16$ & $ 0.3\pm2.1$ & $1.04\pm0.11$ & $
0.4\pm1.5$ 
& $1.03\pm0.13$ & $ 0.4\pm1.8$ & C  \nl 
6838 &  M~71      & $0.86\pm0.09$ & $-1.7\pm1.1$ & $0.81\pm0.06$ &
$-2.4\pm0.8$ 
& $0.84\pm0.07$ & $-2.1\pm1.0$ & Y? \nl 
7078 &  M~15      & $0.94\pm0.07$ & $-0.7\pm1.0$ & $1.02\pm0.04$ & $
0.3\pm0.5$ 
& $0.98\pm0.05$ & $-0.2\pm0.8$ & C  \nl
  -  &  Pal~12    & $0.65\pm0.06$ & $-4.5\pm0.8$ & $0.57\pm0.04$ &
$-5.6\pm0.6$ 
& $0.61\pm0.05$ & $-5.0\pm0.7$ & Y  \nl
\hline
\enddata
\end{deluxetable}

\newpage

\begin{deluxetable}{crrrrrr}
\tablecolumns{10}
\tablewidth{0.0pt}
\tablecaption{
Coefficients of the polynomials used to interpolate the
theoretical quantities listed as column headers.
\label{coeff}}
\tablenum{3}
\tablehead{
\multicolumn{1}{c}{} 		& 
\multicolumn{2}{c}{V99} 	&
\multicolumn{2}{c}{SCL97} 	& 
\multicolumn{2}{c}{C98} 	\\
\multicolumn{1}{c}{Coeff} 		 &
\multicolumn{1}{c}{$M_V$(TO)} 		 &
\multicolumn{1}{c}{$\delta(V-I)_{@2.5}$} & 
\multicolumn{1}{c}{$M_V$(TO)} 		 &
\multicolumn{1}{c}{$\delta(V-I)_{@2.5}$} & 
\multicolumn{1}{c}{$M_V$(TO)} 		 & 
\multicolumn{1}{c}{$\delta(V-I)_{@2.5}$} 	
}
\tiny
\startdata
a       & $ 1.599700$ 	& $ 0.828002$  & $-1.580420$ 	& $ 0.930417$ & $ 
11.99130$ 	& $ 0.667046$  \nl		
b       & $ 0.825196$ 	& $-0.165156$  & $ 1.111200$ 	& $-0.356030$ & $ 
2.550110$ 	& $-0.217992$  \nl
c  	& $ 3.801050$ 	& $-0.812187$  & $ 12.85960$ 	& $-1.472910$ & 
$-23.50860$ 	& $-0.123297$  \nl
d 	& $-0.278009$ 	& $ 0.191788$  & $-0.226336$ 	& $ 0.186956$ & 
$-0.331870$ 	& $ 0.243417$  \nl
e 	& $-1.428870$ 	& $ 0.589075$  & $-10.29670$ 	& $ 1.491670$ & $ 
21.91010$ 	& $-0.008126$  \nl
f 	& $-1.153670$ 	& $ 0.611913$  & $-1.716990$ 	& $ 0.970103$ & 
$-4.612900$ 	& $ 0.857613$  \nl
g 	& $-0.074459$ 	& $ 0.021023$  & $-0.025408$ 	& $ 0.020722$ & 
$-0.061030$ 	& $ 0.034687$  \nl
h 	& $ 0.339905$ 	& $-0.223764$  & $ 3.240170$ 	& $-0.590396$ & 
$-6.108660$ 	& $-0.069297$  \nl
i 	& $-0.007749$ 	& $-0.077650$  & $ 0.103320$ 	& $-0.077396$ & $ 
0.094752$ 	& $-0.080693$  \nl
j 	& $ 0.390382$ 	& $-0.312448$  & $ 0.754886$ 	& $-0.477246$ & $ 
2.124730$ 	& $-0.421044$  \nl
\hline
$rms:$ 	&  0.017 mag	& 0.002 mag    & 0.010 mag	& 0.001 mag   & 0.017 
mag	& 0.001 mag    \nl
	& (=0.23 Gyrs)	& (=0.15 Gyrs) & (=0.12 Gyrs)	& (=0.15 Gyrs)& (=0.20 
Gyrs)	& (=0.12 Gyrs)
\enddata
\end{deluxetable}

\end{document}